\documentclass[aps,prd,showpacs]{revtex4}
\usepackage{graphicx}
\usepackage{epsfig}
\usepackage{psfig}
\usepackage{revsymb}
\begin{document}
\title{The Upper Limit of Magnetic Field Strength in Dense Stellar
Hadronic Matter}
\author{Somenath Chakrabarty}
\affiliation{
Department of Physics, Visva-Bharati, Santiniketan 731 235, 
West Bengal, India, E-mail:somenath@vbphysics.net.in}

\pacs{97.60.Jd, 97.60.-s, 75.25.+z} 
\begin{abstract}
It is shown that in strongly magnetized neutron stars, there exist upper
limits of magnetic field strength, beyond which the self energies for both 
neutron and proton components of neutron star matter become complex in nature. 
As a consequence they decay within the strong interaction time scale. However, 
in the ultra-strong magnetic field case, when the zeroth Landau level is only 
occupied by protons, the system again becomes stable against strong decay.
\end{abstract}
\maketitle
The study of the effect of strong quantizing magnetic fields on dense
neutron star matter has become extremely interesting and also important
after the recent discovery of a few strongly magnetized neutron
stars, known as magnetars \cite{R1,R2,R3,R4}, These strange stellar objects are 
supposed to be relatively young (age $\sim10^4$yrs.) neutron stars and 
are also believed to be the possible sources of anomalous X-ray pulses and 
soft gamma ray emissions (AXP and SGR). If the magnetic fields are strong enough,
particularly at the central region, then most of the physical properties of 
these strange stellar objects and also various physical processes taking place 
at the core region should change significantly. The strong magnetic field at the 
core region affects the equation of state of dense neutron star matter 
\cite{R6,R7}. It is seen that in the case of a compact neutron star, the 
possibility of phase transition to quark matter at the core region is totally 
forbidden if the magnetic field strength exceeds $10^{15}$G \cite{R9}.
The elementary processes, e.g. weak and electromagnetic reactions and decay 
processes are found to be strongly influenced by such 
intense magnetic fields. As the cooling of neutron stars are mainly due to
the emission of neutrinos through weak processes (URCA or modified URCA), 
the presence of strong quantizing magnetic field should significantly affect 
the thermal evolution of neutron stars \cite{R11}. The presence of
strong magnetic field affects chiral properties of both QED and QCD matter and
the corresponding vacuum, in low as well as in $3+1$ dimensions. The strong 
quantizing magnetic field acts like a catalyst to generate the mass dynamically 
\cite{R18,R19,R21,SCH}.

Motivated by a recent study of Shabad and Usav \cite{RSU}, in the present article
we shall investigate the possibility of upper limit of magnetic field strength 
in neutron star matter beyond which the hadronic matter becomes unstable. 
We shall follow some of our recent studies where we have developed a relativistic
formalism of Landau theory of Fermi liquid for dense neutron star matter in 
presence of strong quantizing magnetic field with the exchange of 
$\sigma-\omega-\rho$ mesons \cite{R25,R25a}. We have noticed that the densities 
of both proton and neutron components, which are in $\beta$-equilibrium are 
functions of magnetic field strength. This is a consequence of variation of weak 
interaction rates with magnetic field. In this article we shall follow the 
formalism developed recently in ref. \cite{R25a} to obtain the complex 
self-energies for both neutron and proton of dense neutron star matter. The 
detail calculations are available in the last reference. 

In this study, we have considered the elementary processes as shown in fig.(1). 
In this figure, we are particularly interested in the exchange diagram of $n-p$ 
scattering processes with the transfer of $\sigma-\omega-\rho_3$ mesons and 
direct diagram with $\rho_\pm$ meson exchange. The self-energies of neutron and 
proton components become complex in nature because of these two diagrams. Now the
underlying technique to obtain complex self-energies is to evaluate the
Lanadu-Fermi liquid interaction function (also known as the quasi-particle
interaction function) from two particle forward scattering matrix for these two 
diagrams. The self-energies for proton and neutron components are then
obtained by evaluating the momentum integrals of respective quasi-particle 
interaction functions (here we have assumed that the temperature $T=0$ for
neutron star matter and therefore the upper limits of both proton and neutron 
momentum integrals are the corresponding Fermi momentum). Now it is easy to see 
from reference \cite{R25a} that in the evaluation of proton and neutron self
energies we needed the following traces of projection operators:\par
1. For $\sigma$-meson exchange case, we need
${\rm{Tr}}[\Lambda_{pn}(x,p)\Lambda_{pn}(x^\prime,p^\prime)]$.
Which is found to be complex in nature.\par
2. In the case of $\omega$-meson exchange, it is necessary to evaluate the trace
${\rm{Tr}}[\Lambda_{pn}(x,p)\gamma^\mu\Lambda_{pn}(x^\prime,p^\prime)\gamma_\mu]$.
It is seen that only $\mu=0$ and $z$ give non-zero contributions and both of 
them are complex in nature.\par
3. In the case of neutral $\rho$-meson exchange, the contribution is exactly
same as that of $\omega$-meson case, except a factor of $1/2$.\par
4. Finally in the case of charged $\rho$ meson exchange ($n-p$ direct scattering
diagram) we have to evaluate ${\rm{Tr}}[\Lambda_{pn}(x,p)\gamma^\mu]
{\rm{Tr}}[\Lambda_{pn}(x^\prime,p^\prime)\gamma_\mu]$.
In this case also only for $\mu=0$ and $z$, the traces are non-zero and
are found to be again complex in nature.\par
The projection operators are given by
\begin{equation}
\Lambda_{np}(x,p)=u_n^\uparrow \bar u_p^\uparrow +
u_n^\downarrow \bar u_p^\downarrow
\end{equation}
and
\begin{equation}
\Lambda_{pn}(x,p)=u_p^\uparrow \bar u_n^\uparrow +
u_p^\downarrow \bar u_n^\downarrow
\end{equation}
with 
\begin{equation} 
u^{\uparrow}=\frac{1}{[E_{\nu}(E_{\nu}+m)]^{1/2}}
\left( \begin{array}{c}
(E_{\nu}+m)I_{\nu;p_y}(x)\\0\\p_zI_{\nu;p_y}(x)\\
-i(2\nu q_iB_m)^{1/2}I_{\nu-1;p_y}(x)
\end{array} \right)  
\end{equation}
and
\begin{eqnarray}
u^{\downarrow}=\frac{1}{[E_{\nu}(E_{\nu}+m)]^{1/2}}
\left( \begin{array}{c}
0\\(E_{\nu}+m)I_{\nu-1;p_y}(x)\\i(2\nu q_iB_m)^{1/2}I_{\nu;p_y}(x)\\
-p_zI_{\nu -1;p_y}(x)
\end{array} \right) 
\end{eqnarray}
are the Dirac spinors, with the symbols $\uparrow$ and $\downarrow$ for up and 
down spin states respectively and
\begin{eqnarray}
I_\nu=\left (\frac{qB_m}{\pi}\right )^{1/4}\frac{1}{(\nu !)^{1/2}}2^{-\nu/2}
\exp \left [{-\frac{1}{2}qB_m\left (x-\frac{p_y}{qB_m} \right )^2}\right 
]\nonumber \\
H_\nu \left [(qB_m)^{1/2}\left (x-\frac{p_y}{qB_m} \right) \right ]
\end{eqnarray}
with $H_\nu$ the well-known Hermite polynomial of order $\nu$ and
$L_y,~~L_z$ are length scales along $y$ and $z$ directions respectively.
Here $\nu$ is the Landau quantum number proton proton and can have any integer 
value including zero. For neutron as neutral particle, we have used the standard 
Dirac spinor solutions. As we have found in ref.\cite{R25a} that with 
these spinor solutions, the traces as mentioned before become complex in nature 
for the diagrams we have considered.

Here we have chosen the gauge $A^\mu \equiv(0,0,xB_m,0)$, so that the constant 
magnetic field $B_m$ is along $z$-axis. Now the Landau levels for the protons 
will be populated if the magnetic field strength $B_m$ exceeds 
the quantum critical value $B_m^{(c)(p)}=m^2_p/q_p \approx 1.6\times 10^{20}$G, 
where $m_p$ and $q_p$ are proton mass and charge respectively. 
In the relativistic region the quantum critical value is the typical
strength of the magnetic field at which the proton cyclotron quantum exceeds the 
corresponding rest mass energy or equivalently the de Broglie wave length for 
proton exceeds the corresponding Larmor radius. Now for $B_m >B_m^{(c)(p)}$, 
with the relation  $p_F^2 \geq 0$, the maximum value of Landau quantum number at 
$T=0$ is given by
\begin{eqnarray}
[\nu_{max}^{(p)}]=\frac{(\mu_{p}^{2}-m_{p}^{2})}{2q_{p}B_{m}}
\end{eqnarray} 
which is an integer but less than the actual value of right hand side and $\mu_p$
is the proton chemical potential. The external magnetic field will therefore 
behave like a classical entity if the strength is less than the quantum 
threshold value and in this region one has to use the plane wave solution with 
standard form of Dirac spinors for both proton and neutron components. Then it 
is possible to obtain the equation of state of dense neutron star matter in a 
straight forward manner following reference \cite{BC}. Further, it is obvious 
from eqn.(6) that for a given proton density, the upper limit for $\nu$ can
become zero for some large magnetic field strength. In that case only the
zeroth Landau level will be occupied by protons. For such strong magnetic
field since only allowed value of Landau quantum number for proton is
$\nu=0$, then using the relation that $H_{-n}=0$ for integer $n>0$, 
the spin up component (eqn.(3)) reduces to
\begin{equation} 
u^{\uparrow}=\frac{1}{[E_{0}(E_{0}+m)]^{1/2}}
\left( \begin{array}{c}
(E_{0}+m)I_{0;p_y}(x)\\0\\p_zI_{0;p_y}(x)\\
0
\end{array} \right)  
\end{equation}
whereas the down spin component  (eqn.(4)) becomes a null column matrix. It's 
physical meaning is that for such a strong magnetic field all the protons will
come down to the zeroth Landau level with the spins aligned along the direction of
magnetic field. The crucial point which should be noted here is that in this case
if one evaluates the traces as mentioned before they are no longer be complex
in nature, they will be absolutely real quantities, which actually means that the 
self-energies will be real for both proton and neutron components. To explain 
the point a little more detail, we have plotted in fig.(2) the density of proton 
as a function of quantum critical limit for magnetic field strength. Since the 
neutron star matter is in $\beta$-equilibrium and rates of the weak interaction 
processes (URCA and modified URCA) are strongly dependent on the strength of 
magnetic field, the equilibrium density of proton will be a function of magnetic 
field strength. Which means that the proton or neutron density can not be
arbitrary for $B_m>B_m^{(c)(p)}$. 

Let us now consider fig.(2) and try to understand different regions as indicated 
in the diagram. At the bottom, since $B_m<B_m^{(c)(p)}$, it is the classical 
region. The Landau levels for protons are not populated in this region, they are 
represented by plane wave solutions with standard Dirac spinors. As a result the 
dense neutron star matter will be stable in this region. This region is marked 
by {\it{stable classical region}}.

Next we consider the other stable region at the top of this figure. In
this region the strength of magnetic field is such that the maximum value
of Landau quantum number of protons is zero. The Dirac spinor for proton is
given by eqn.(7), which makes the self energies absolutely real. As a
consequence the matter is again stable against strong decay. Since the
quantum mechanical effect of strong magnetic field is very much important
in this region, we call it as {\it{stable quantum region}}.

We now consider the region which in our opinion is the most important part
of this figure and the aim of this article is to show that for dense
stellar hadronic matter such a region in density-magnetic field space actually 
exists in magnetized neutron stars. Within this region the spinor solutions for 
proton are given by eqns.(3) and (4) and as a consequence the self energies of 
both proton and neutron components are complex in nature. Which makes the system 
in this region unstable. Both proton and neutron decay within the strong 
interaction time scale as shown in fig.(3). This region started from a point 
with $B_m=B_m^{(c)(p)}$ and the proton density at this point is the lowest 
allowed value inside a magnetized neutron star with quantum critical strength 
of magnetic field. Since the energy of the system decreases as more and more 
protons occupy the low lying Landau levels, the density of proton increases with 
the increase of magnetic field. It occurs within the neutron star by the 
conversion of more and more neutrons to protons through weak processes in 
presence of strong magnetic field.

Finally, the region at the top right corner is forbidden both classically and
quantum mechanically. Since in this region the magnetic field strength is greater 
than the quantum critical value, this part can not be treated as classical.
Whereas, quantum mechanically the density of protons in $\beta$-equilibrium can 
not be so low. Therefore the hadronic matter within strongly magnetized neutron 
stars can not occupy this region, We have marked it by {\it{forbidden region}}.

The horizontal line, separating the stable classical zone and the unstable
region is for $B_m=B_m^{(c)(p)}$ against $n_p$.
The diagonal line separating the upper stable region and the unstable
region is for $[\nu_{\rm{max}}]=1$ for protons. The stable region above
this line is for $[\nu_{\rm{max}}]=0$, whereas within the unstable region, 
$[\nu_{\rm{max}}]$ can have all possible integer values beginning with unity.

Hence we finally conclude that in a strongly magnetized neutron star,
if the magnetic field strength is $>1.6\times 10^{20}$G, the quantum critical 
value for protons, the matter can not be stable and therefore this is the upper 
limit for a neutron star magnetic field with stable hadronic matter.

\newpage
\begin{figure} 
\psfig{figure=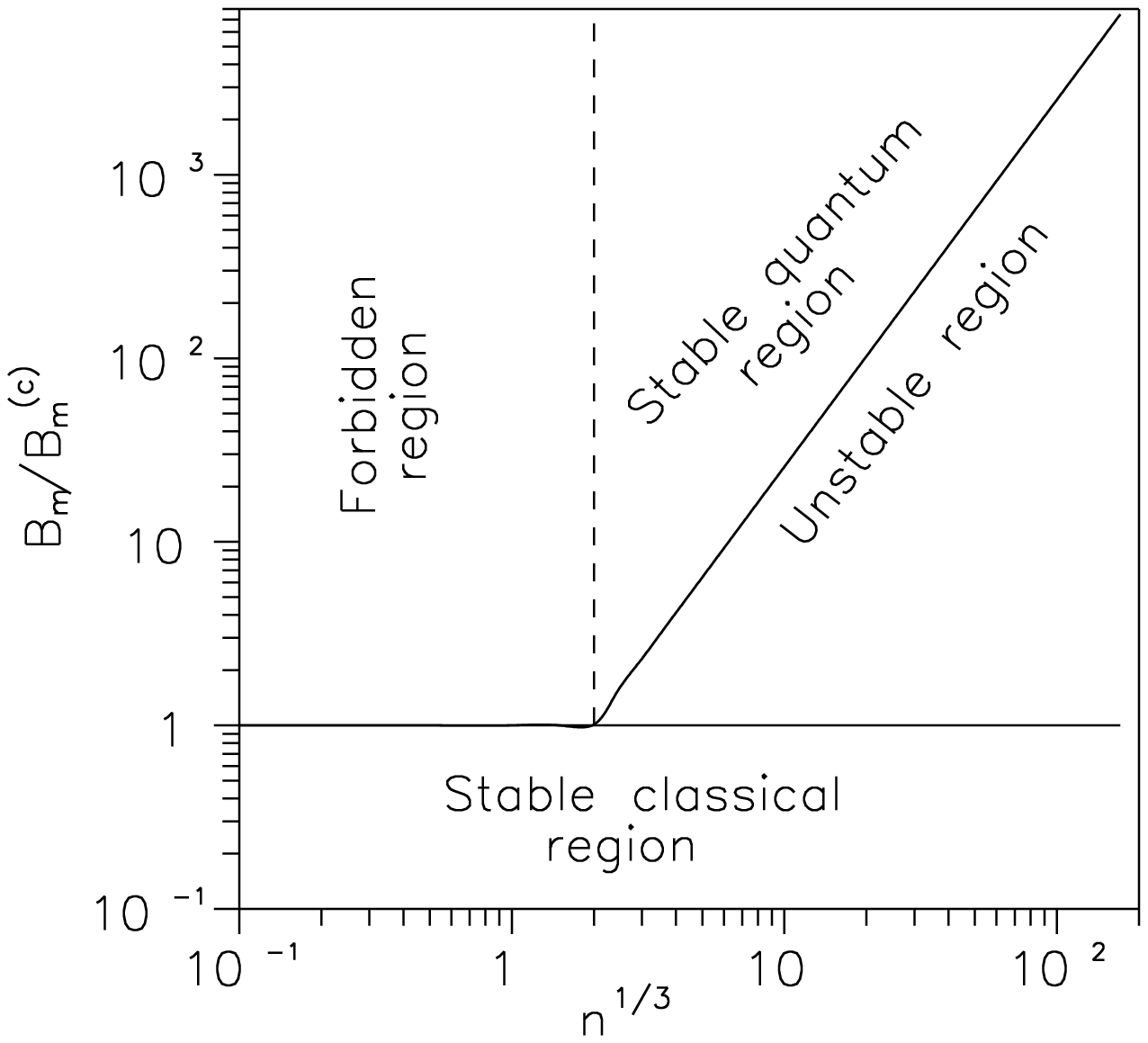,height=0.5\linewidth}
\caption{Magnetic field (expressed as $B_m/B_m^{(c)}$) plotted against the
density of proton matter (in fm$^{-1}$)}
\end{figure}
\begin{figure} 
\psfig{figure=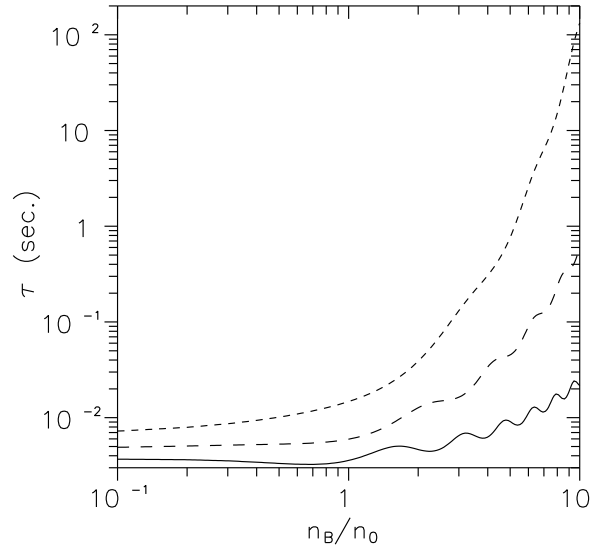,height=0.4\linewidth}
\caption{Variation of proton and neutron life times (in sec. in unit of
$10^{22}$)  with proton
density (expressed in terms of normal nuclear density $n_0=0.17$fm$^{-3}$).
The upper curve is for $B_m=10^{18}$G, middle one is for $B_m=10^{16}$G and
the lower one is for $B_m=10^{14}$G.}
\end{figure}
\end{document}